# Dynamical scaling and isotope effect in temporal evolution of mesoscopic structure during hydration of cement

S. Mazumder[1*], D. Sen[1], R. Loidl[2] and H. Rauch[2]

[1]*Solid State Physics Division, Bhabha Atomic Research Centre, Mumbai 400 085, India;* [2]*Atominstitut der Österreichischen Universitäten, A-1020 Wien, Austria and Institut Laue-Langevin, BP 156, 38042 Grenoble Cedex 9, France*

* To whom correspondence should be addressed, E-mail: smazu@barc.gov.in

**Abstract**

The evolution of mesoscopic structure for cement-water mixtures turning into colloidal gels remains far from being understood. Recent neutron scattering investigations (Phys. Rev. Lett. 93, 255704 (2004); Phys. Rev. B. 72, 224208 (2005); Phys. Rev. B. 82, 064203 (2010)),, reveal the role of hydrogen bond in temporal evolution of the mesoscopic structure during hydration of cement which is the most consumed synthetic material. The present neutron scattering investigation on hydration of cement with a mixture of light and heavy water points to incomprehensibility of the temporal evolution of the mesoscopic structure in terms of earlier observations on hydration with pure light or heavy water.  Unlike in the case of hydration with light water, disagreement has been observed with the hypothesis of dynamical scaling for hydration of cement with a mixture of the two types of water. The dynamics of evolution of the mesoscopic structure has been observed to be nonlinear in regard to the composition of hydration medium.

## I. INTRODUCTION

Investigations on dynamics of new phase formation involve mapping of time-dependent scattering function $S(\mathbf{q},t)$ where t stands for time and $q$ is the modulus of the scattering vector $\mathbf{q}$.  Because of isotropic nature of the system at mesoscopic scale, $S(\mathbf{q},t)$ is only a function of $q$. At late stages, the dynamics of new phase formation is highly nonlinear process far from equilibrium. The new phase forming systems exhibit self-similar growth pattern with dilation symmetry, with time dependent scale, and scaling phenomenon[1]. The scaling hypothesis assumes the existence of a single characteristic length scale L(t), such that the domain sizes and their spatial correlation are time invariant, as depicted in Fig. 1, when the lengths are scaled by L(t). Exhibition of dynamical scaling implies that domains, very large in number for a conceivable macroscopic system, are in communication with each other such that they are scaled by the same characteristic length L(t). For a d-dimensional Euclidean system, simple scaling ansatz[1], S(q,t)=L(t)$^d$ $\tilde{F}$(qL(t)), holds good at a later time, where $\tilde{F}$ (x) is the scaled scattering function.

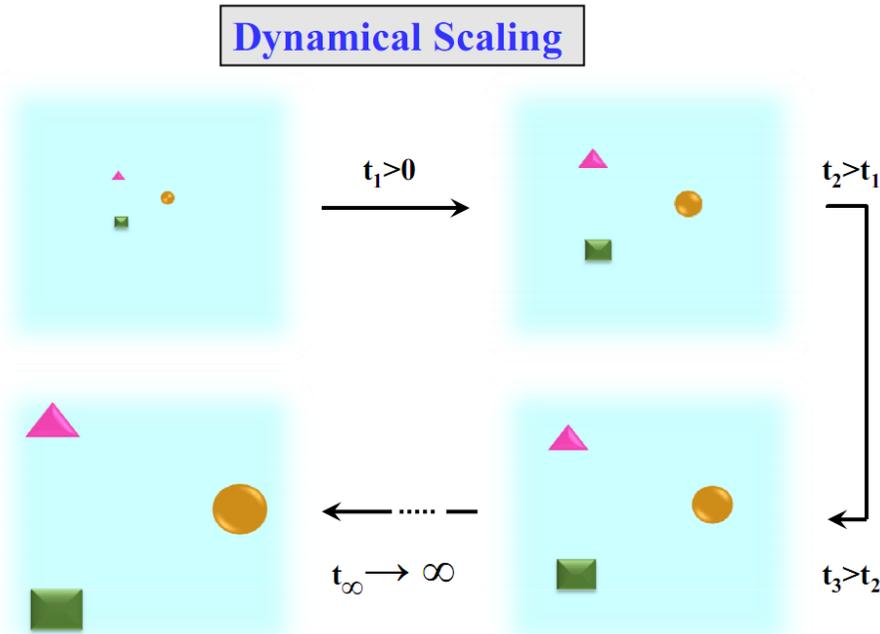

Fig. 1. *A schematic representation of dynamical scaling. Individual inhomogeneities are scaled up by same time dependent scale factor L(t) implying some communication mechanism between the individual inhomogeneities.*

Reaction of cement with water is exothermic in nature with production of crystalline Ca(OH)$_2$ and non-crystalline colloidal gel like calcium silicate hydrate (C-S-H). The progress of the reaction, as manifested by the heat evolution rate, has been observed[2-4] to be oscillatory in nature indicating different stages of hydration. It is to be noted that some structural changes in the system can also cause liberation or absorption of heat. However, compressive strength of the cement-water mixture increases[5] monotonically with time albeit with varying slopes. The development of compressive strength is also indicative of existence of one or more processes like chemical, structural and combination of both in the system. It is desirable to know the contribution and relative significance of each of these processes towards development of compressive strength. It is only expected that temporal evolution of compressive strength bears some relation, hitherto unknown, with temporal change of heat liberation rate.

Progress of the reaction, as monitored by temporal evolution of S(q,t) in recent neutron scattering measurements[6-9], has been observed to be non-linear in nature strongly dependent on the scale of observation and on the medium of hydration (light or heavy water). Although investigation[10] on cement is more than a century old, the temporal evolution of mesoscopic structure, as revealed by these measurements, does not exhibit a well defined correlation with the existing data on time dependent change in either compressive strength or heat liberation rate.

The mesoscopic structure of C-S-H gel determines the desirable properties of hardened cement. To elucidate the microscopic structure of the C-S-H gel, many models[11-16] have been proposed. The investigations[6-9,17,18], based on small-angle neutron scattering (SANS), on continuous temporal evolution of mesoscopic structure during hydration of cement are only recent. Small-angle scattering (SAS) technique can elucidate mesoscopic structure in the length scale of $10^2 - 10^4$ Å and is an ideal probe for time-resolved structural investigation on C-S-H gel. Both neutrons and X-rays are used as probing radiation for SAS investigations. Neutrons, owing to their considerably higher penetration power, probe larger sample volume. Neutrons are not invasive as they leave the structure of the sample unaltered, and scatter very effectively from hydrogen-bearing materials such as those found in hydrated cements.

Small-angle X-ray[19] scattering (SAXS) and Small-angle neutron[20] scattering (SANS) studies have demonstrated the fractal morphology of C-S-H gel. Further, SAXS investigations have demonstrated[19,21] that C-S-H gel undergoes transition from a ramified to a relatively more compact structure. A structure is said to be more ramified[22] at a particular point if more number of bonds are to be eliminated to isolate an arbitrarily large bounded sub-structure surrounding that point. All other structural parameters remaining same, a more ramified structure will have lesser mass and hence more open vis-à-vis one with lesser ramified structure. For a ramified mass fractal structure with fractal dimension $D_m$ embedded in a three-dimensional space, the value of $D_m$ lies in the range $1<D_m<3$. For a ramified rod and disk, the value of $D_m$ lie in the range $0<D_m<1$ and $1<D_m<2$, respectively. The smaller the value of $D_m$, the more ramified is the object.

Examinations, involving continuous monitoring of temporal evolution of mesoscopic structure during hydration of cement with light water ($H_2O$) and heavy water ($D_2O$), have been reported[6-9] only recently. It has been observed[6-9] that the kinetics, of hydration of silicates and sulphates with light and heavy water, are of nonlinear nature even at the initial time. While the formation of hydration products is synchronous for hydration with $H_2O$, the process is non-synchronous for hydration with $D_2O$. Time-resolved experiments[6-7], with hydration time not exceeding five hours, to investigate the temporal evolution of morphological features, in the length scale $10^3 - 10^4$ Å, of hydrated gel C-S-H and C-S-D have indicated that for hydration of cement with $H_2O$, topographical mesoscopic structure could not be described in terms of classical porous medium with a well-defined specific inner surface. The mesoscopic structure, in the length scale $10^3 - 10^4$ Å, is that of a mass fractal initially.

In the case of hydration of silicates, with light water, the hydrating mass exhibits[6-7] a mass fractal nature, in the length scale $10^3 - 10^4$ Å, for the initial few (2-3) hours of hydration, with the mass fractal dimension increasing with time and reaching a plateau after about 150 min with the maximum attained value less than 3. The second phase grows with time initially. Subsequently, the domain size of the second phase saturates. Temporal evolution of square of the linear dimension of the inhomogeneity mimics[6-7] the trend of the temporal evolution of the fractal dimension. In the large time limit, the temporally evolving system exhibits self-similar growth pattern with dilation symmetry and with the scaling phenomenon[1].

In view of aforementioned observations[6-9], it has been concluded that for hydration of cement with $H_2O$, water rich mass fractal C-S-H sol and crystalline $Ca(OH)_2$, are formed initially. $Ca(OH)_2$ phase is a minor phase and the fractal structure as observed in scattering measurements pertains to major C-S-H phase as pores cannot exhibit mass fractal nature. Both the fractal dimension and linear dimension of mass fractal C-S-H phase grow with time. The initial increase of mass fractal dimension with time reflects the transition from a ramified and porous structure to a relatively more compact homogeneous solid matrix. As the system exhibits dynamical scaling phenomenon, ratio of linear sizes of $Ca(OH)_2$ crystallite and C-S-H phase must be constant indicating the growth, Fig. 1, of linear sizes of both the phases by the same characteristic length with time implying the synchronous formation of $Ca(OH)_2$ and C-S-H in this period of hydration. In the length scale $10^2 - 10^3$ Å, morphological change[9] of the mesoscopic structure is much more monotonic in nature.

The hydration of silicates with light or heavy water is expected to be quite similar except for their kinetics. Due to higher molecular mass, diffusion is expected to be more sluggish for heavy water. Further, it is known that the hydrogen bond with deuterium is slightly stronger than the one involving ordinary hydrogen[23]. This is due to higher reduced mass and hence, lower energy at the same level including zero-point energy of the bond of diatomic molecule involving deuterium vis-à-vis that involving hydrogen – responsible also for shorter bond length in O-D vis-à-vis that for in O-H. With a lower energy, more energy is required to overcome the activation barrier for bond cleavage or dissociation. In fact, with heavier elements like oxygen, calcium or silicon, the frequency or energy of a bond involving D is approximately $1/\sqrt{2}$ times that of the corresponding bond involving H. Hence, it is expected that hydration with $D_2O$ will be slower as compared to that with $H_2O$ because of higher activation energy. Moreover, hydration products involving deuterium will be more stable than those involving hydrogen. The lifetime of hydrogen bond involving D is longer vis-à-vis that involving H because the libration motions perpendicular to the bond direction have smaller amplitude for D than that for H, because of difference in isotopic masses.

Some contrasting behavior has been observed[6-7,9] in the case of hydration of silicates with heavy water. The domain size of the density fluctuations grows in the beginning for a while, and subsequently shrink with time. Mesoscopic structure, in the length scale $10^3 - 10^4$ Å, undergoes transition from mass fractal to surface fractal and finally again to mass fractal. No agreement has been observed[6-7] with the dynamical scaling hypothesis with all possible measures of the characteristic length. In the length scale $10^2 - 10^3$ Å , morphological change[9] of the mesoscopic structure is much wider. It is a conjecture that the different rates of diffusion of light and heavy water in forming a gel structure in silicates lead to the formation of different structural networks with different scattering contrasts.

The above observations call for examining the temporal evolution of the mesoscopic structure encountered during hydration of cement with mixture of $H_2O$ and $D_2O$. Investigations on non-linearity and its modulations when the hydration medium is mixed is a subject of importance. This also to understand better the role of hydrogen bond in the hydration behavior of cement

## II. EXPERIMENT

Water mixtures, comprising of light water ($H_2O$) and heavy water ($D_2O$) in the molar ratio of 3:1, 1:1 and 1:3, were prepared initially for hydration of cement. Pure powder specimens of $C_3S$ were mixed with water mixtures at varying water-to-cement mass ratio (w/c), ranging from 0.3 to 0.5 to obtain a wet mass. C-S-H gel is known[6-7,9] to have fractal mesoscopic structure in the length scale $10^2 – 10^3$ Å. A preliminary measurement with a medium resolution SANS facility[24] indicated that hydrated cement has a fractal microstructure on the length scale 1-100 nm.. It also indicated the possibility of existence of inhomogeneities larger than 100 nm. So SANS measurements were carried out with Ultra Small Angle Neutron Scattering instrument[25] S18 at 58 MW high flux reactor at ILL, France. The wavelength $\lambda$ used was 1.87 Å. The scattered intensities were recorded as a function of q ($=4\pi (\sin \theta)/\lambda$, $2\theta$ being the scattering angle). The scattering data were corrected for background and primary beam geometry. For background, scattering measurements have been carried out at about $q=10^{-2}$ $Å^{o-1}$. At the outset, it was inferred that hydrating specimens are isotropic in nature and so is the scattering function[26] $S(q,t)$.

## III. DATA INTERPRETATION AND DISCUSSION

Present work reports the mesoscopic structural investigations on real time hydration of cement. Widely used Portland cement[5] is a composite material consisting of fine crystalline grains of tri-calcium silicate, $3CaO.SiO_2$ (Abbreviation $C_3S$; Approx. mass percentage range 60-80%) along with minor constituents like di-calcium silicate, tri-calcium aluminate, tetra-calcium iron aluminate etc. As the major constituent of Portland cement, $C_3S$ can be used as a model for hydration of cement. The hydration reaction of $C_3S$ can be written as

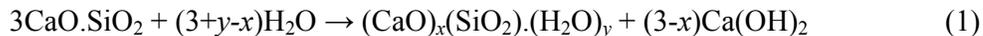

$$3CaO.SiO_2 + (3+y-x)H_2O \rightarrow (CaO)_x(SiO_2).(H_2O)_y + (3-x)Ca(OH)_2 \qquad (1)$$

where x and y vary, $x$ is bounded by $0 \leq x \leq 3$. y, measures the partitioning of reacted $H_2O$ between $(CaO)_x(SiO_2).(H_2O)_y$ and $Ca(OH)_2$ phases, is bounded on one side i.e. $y \geq 0$. $y=0$ indicates the consumption of entire reacted $H_2O$ towards formation of $Ca(OH)_2$. $x=3$ indicates hydration of $C_3S$ without formation of $Ca(OH)_2$. That $x$ is time dependent and functional form of $x(t)$ is hydration medium ($H_2O$ or $D_2O$) dependent have been established[9] only recently. pH dependence of the kinetics of the reaction is a plausible reason for the time dependence of x(t). The product $(CaO)_x(SiO_2).(H_2O)_y$ (abbreviated as C-S-H, wherein hyphens indicate variable stoichiometry) is calcium silicate hydrate – a colloidal gel-like polymeric material at the late stage of hydration. Tri-calcium silicate, $3CaO.SiO_2$ (Abbreviation $C_3S$) and $Ca(OH)_2$ are both crystalline in nature. Although, the reaction (1) should be termed as hydrolysis rather than hydration as it involves cleavage of H-OH bond and formation of new bonds, we will continue with the terminology hydration for reaction (1).

Figure 2 depicts the time evolution of scattering function S(q,t) in absolute scale for water mixture hydrating $C_3S$ with w/c = 0.3 and w/c = 0.5, respectively. Different curves in each frame depict the variation of S(q,t) with q for different recorded times. The inset of the figure depicts temporal variation of S(q,t) in arbitrary scale for three specific values of q as mentioned therein. For a system of polydisperse population of inhomogeneities, with degree of polydispersity (m), S(q,t) is given by

$$S(q,t) = \sum_{i=1}^{m} \tau_i(t) V_i^2(t) \rho_i^2(t) P_i(q) = \sum_{i=1}^{m} \tau_i(t) \zeta_i^2(t) P_i(q)$$

where the subscripted quantities $\tau_i$, $V_i$, $\rho_i$, $P_i(q)$ and $\zeta_i$ (= $V_i\rho_i$) are respectively the number density, volume, scattering length density, normalized scattering form factor and total scattering length integrated over volume of the $i^{th}$ type of inhomogeneity. Normalized scattering form factor $P_i(q) \rightarrow 1$ as $q \rightarrow 0$. For a spherical inhomogeneity of radius $r_0$,

$$P(q) = 9 \cdot \frac{(\sin(qr_0) - qr_0 \cos(qr_0))^2}{(qr_0)^6}$$

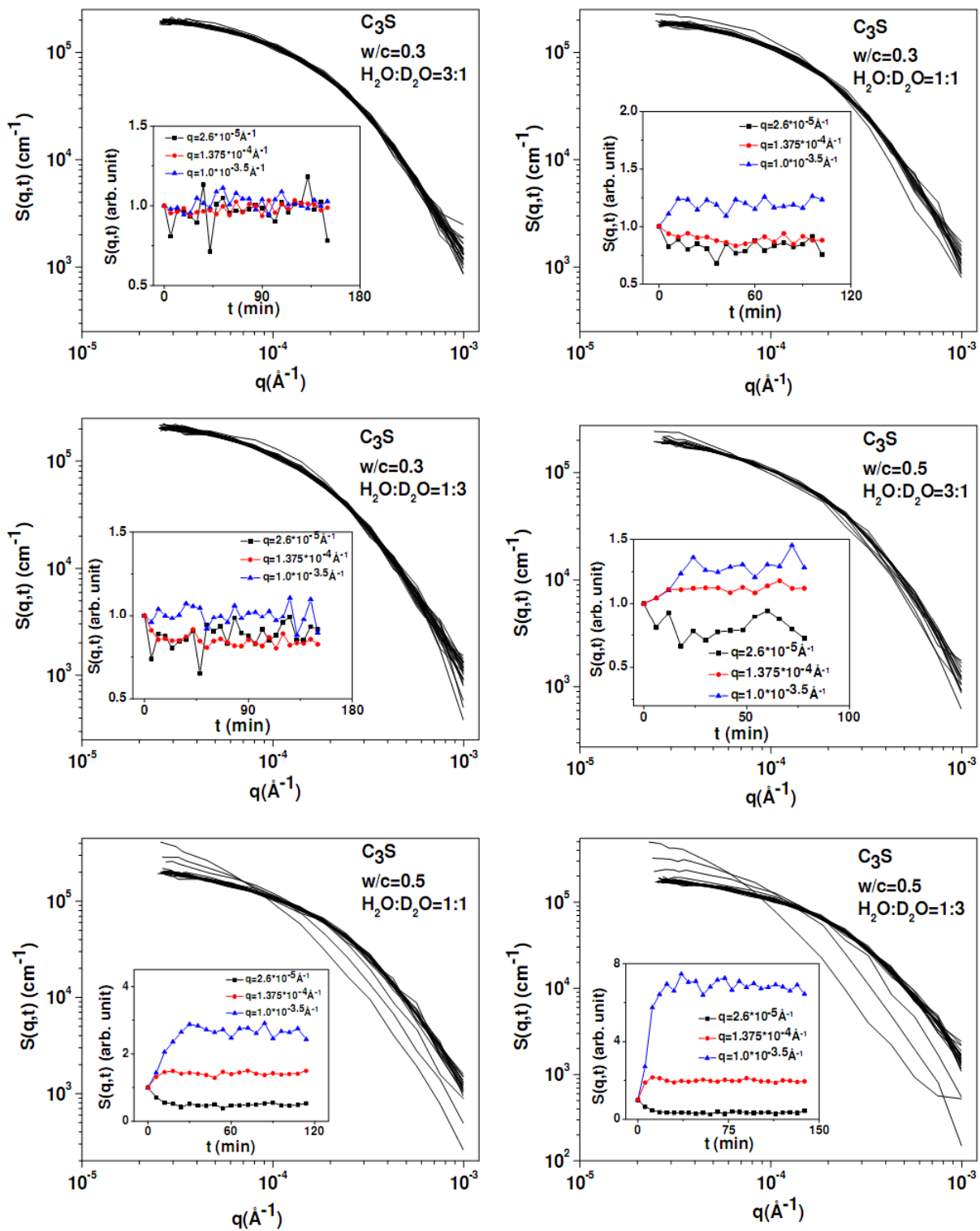

*Fig.2. Time evolution of scattering function $S(q,t)$ for water mixture hydrating $C_3S$ with water-to-cement mass ratio (w/c) = 0.3 and 0.5, respectively. The inset shows the time evolution of $S(q,t)$ at some specified values of q as specified therein. Statistical error bars are smaller than the respective symbol sizes.*

An inhomogeneity is defined by the uniform scattering length density over its volume. It is important to note that S(q,t) has parabolic dependence on $\zeta$.

The oscillatory nature of temporal evolution of S(q,t) for specified q values is indicative of the fact that there are competing factors[9] for temporal evolution of S(q,t) – some causing decay and others causing growth of S(q,t) with time resulting in non-monotonic temporal evolution of S(q,t). To appreciate the oscillatory nature of S(q,t) as depicted in inset of Fig. 2, some model calculations (Model Calculations in Appendix) for hydration of cement have been considered. Some more general cases depicting the oscillatory variation of S(q,t) with t for q=0 has already been dealt with earlier[9].

Insets of Fig. 3 indicate that with increasing hydration time, the curvature of the scattering profiles in the vicinity of $q \rightarrow 0$, varies non-linearly indicating the non-linearity in the growth of pores. The curvature $\kappa(t)$ of normalized S(q,t)) at q is given by

$$\kappa(t) = \frac{|d^2[S(q,t)/S(0,t)]/dq^2|}{(1 + \{d[S(q,t)/S(0,t)]/dq\}^2)^{3/2}} \quad .$$

For monodisperse population of inhomogeneity, the linear dimension of the inhomogeneity is proportional[27] to $\sqrt{\kappa(t)}$. The curvature[27] $\kappa(t)$ in the vicinity of $q \rightarrow 0$ of a scattering profile S(q,t)) is related to

$$G(t) = -d[\ln[S(q,t)/S(0,t)]]/dq^2$$

where G(t) is the negative gradient of the Guinier plot of the normalised scattering profile. For a single scattering profile from monodisperse population of spheres of radius R, in the vicinity of $q \rightarrow 0$, $\kappa = 2R^2/5$ whereas $G = R^2/5$. In subsequent discussions, $\kappa$ is defined in the vicinity of $q \rightarrow 0$ only throughout, if not mentioned otherwise. For a polydisperse population of spherical scatterers, with number density $\rho(R)$ for scatterers of radius R, having the same scattering length density difference, $\kappa$

$=2<R^8>/5<R^6>$ where, $<R^n>$ is the n-th moment of the distribution $\rho(R)$. Curvature[27] and radius of curvature are reciprocal to each other.

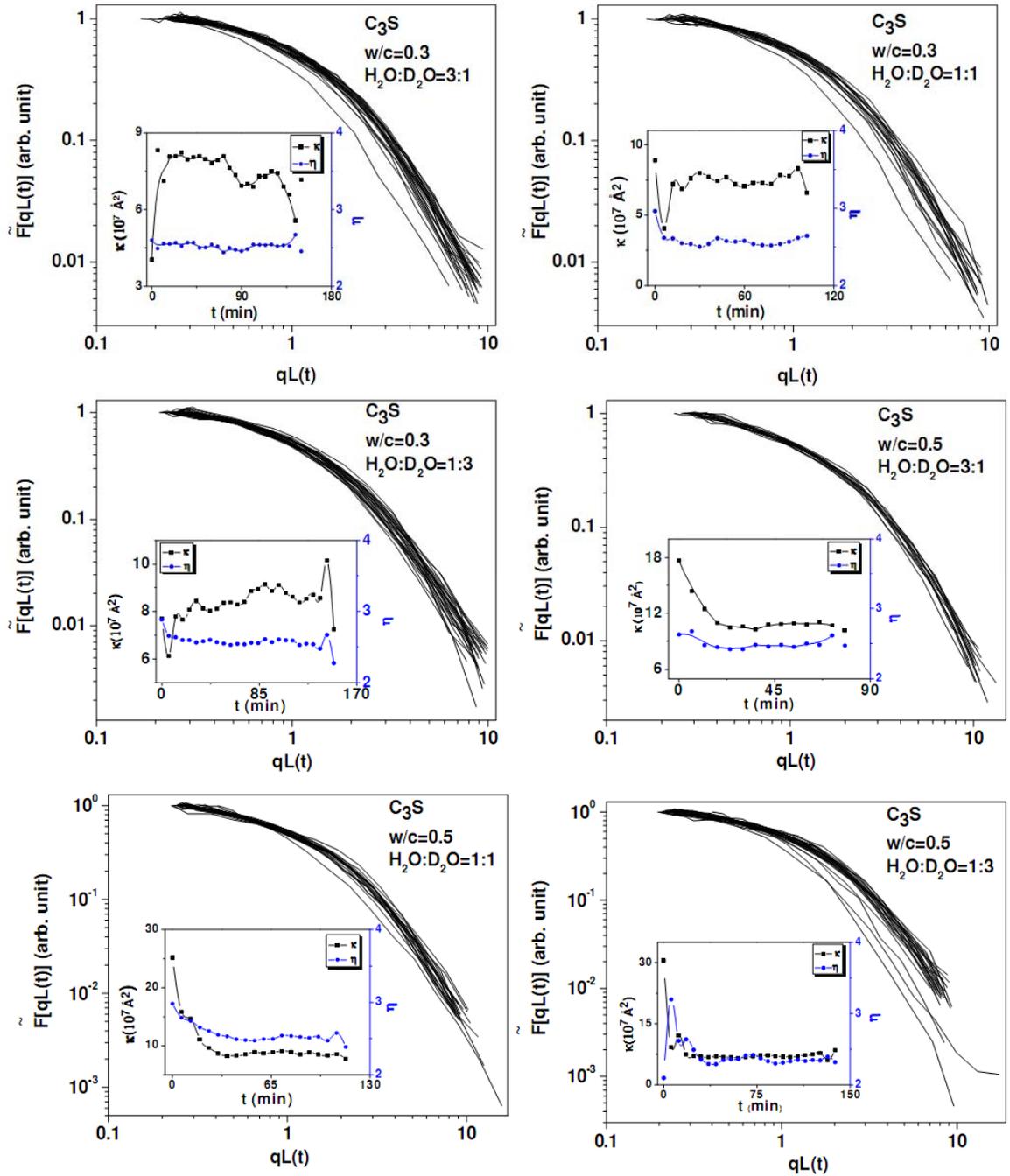

Fig. 3. Scaled scattering function $\tilde{F}[qL(t)]$ for water mixture hydrating $C_3S$ with water-to-cement mass ratio (w/c) = 0.3 and 0.5, respectively. The inset shows the time evolution of $\kappa(t)$ and $\eta(t)$ - associated with power-law scattering. Statistical error bars are smaller than the respective symbol sizes. The solid lines are only guides to the eye.

The insets of Fig. 3 also depict the temporal evolution of $\kappa(t)$ for water mixture hydrating $C_3S$ with w/c = 0.3 and w/c = 0.5, respectively. For all hydrating specimen with w/c = 0.3, $\kappa(t)$ vs. t has oscillatory nature irrespective of mixture composition. For hydrating $C_3S$ with water mixtures, having molar ratios $H_2O : D_2O$ = 3:1 and 1:3, oscillations depict decreasing and increasing trend, respectively, of $\kappa(t)$ vs. t. However, for hydrating specimen with water mixture molar ratio $H_2O : D_2O$ = 1:1, oscillations without any distinct trend are observed.

For hydrating specimen with w/c = 0.5, irrespective of mixture composition, $\kappa(t)$ sharply decreases with time and reaches a plateau. These results are strikingly different from those observed[6-7] for hydration of silicates with pure light water or heavy water. It is pertinent to recall[6-7] that for hydration of silicates with light water, $\kappa(t)$ vs. t has similar generic variation irrespective of value of w/c and so is the case for hydration of silicates with heavy water. For hydration of silicates with light water, $\kappa(t)$ increases with time initially and saturates subsequently. For hydration of silicates with heavy water, $\kappa(t)$ increases[6-7] with time initially, reaches a peak, and finally decreases as time increases.

As recently observed[9] that on mixing silicates with water, water rich C-S-H colloidal sol, a colloidal phase of solid dispersed in continuous liquid phase, particles are formed initially. Depending upon the relative strength of thermal energy and energy barrier between the particles, the aggregation of colloidal particles occurs according to well defined[28] regimes of diffusion-limited cluster aggregation (DLCA) and reaction-limited cluster aggregation (RLCA). DLCA occurs when thermal energy is relatively higher than the barrier energy so that every collision, limited only by the rate of diffusion of the colliding particles, results in sticking the colliding particles together leading to rapid aggregation and cluster mass growing linearly with time. RLCA occurs when thermal energy is relatively lower or comparable to the barrier energy so that out of several collisions one collision results in sticking the colliding particles together leading to slower aggregation and cluster mass growing exponentially with time. In the latter case, aggregation is limited by the probability of overcoming the energy barrier and hence, the name reaction-limited cluster aggregation. In RLCA, cluster with larger mass has more potential bonding sites and grows faster than the smaller ones.

For hydration of silicates with pure light water and heavy water, initial increase[6-7] of $\kappa(t)$ with time is definite not exponential and can be approximated as linear indicating the role of diffusion limited aggregation process in growth of domains. The observed deviation from DLCA regime at the later stage and also somewhat at the initial stage indicate the role of some other effect like hydrogen bond. As hydrogen bond involving deuterium is somewhat stronger than that with hydrogen, it leads to consolidation of the C-S-D mass fractal colloidal particle into a mass with Euclidean core and surface fractal morphology. For hydration of silicates with water mixture, the observed temporal variation of $\kappa(t)$ with t only indicate a plausible role of hydrogen bond.

The insets of Fig. 3 also depict time evolution of Porod exponent $\eta(t)$, as estimated from $\ln(S(q,t))$ vs. $\ln(q)$ in the q range 0.00025-0.001 Å$^{-1}$, for water mixture hydrating $C_3S$ with w/c = 0.3 and w/c = 0.5, respectively. Porod exponent for all the hydrating specimens lies in the range of 2-3 indicating the mass fractal nature of the hydrating paste. For objects whose volume or mass is fractal (cluster aggregates), $S(q,t)$ asymptotically approaches a form $S(q,t) \sim q^{-\eta(t)}$ where the exponent $\eta$ reflects[29] directly the mass fractal dimension $D_m$. For a mass fractal object embedded in a three-dimensional space, $\eta = D_m$ with $1 < \eta, D_m < 3$. For water mixture hydrating specimen, irrespective of mixture composition, $\eta(t)$ decreases marginally initially with time and reaches a plateau with minor oscillations without any distinct trend. The generic variations of $\kappa(t)$ vs. t and $\eta(t)$ vs. t differ widely. These results are strikingly different from those observed[6-7] for hydration of silicates with pure light water or heavy water. It is noteworthy[6-7] that for hydration of silicates with light water, $\eta(t)$ vs. t has similar generic variation as $\kappa(t)$ vs. t irrespective of w/c value and so is the case for hydration of silicates with heavy water. For hydration of silicates with light water, $\eta(t)$ increases[6-7] with time initially and saturates subsequently. For hydration of silicates with heavy water, $\eta(t)$ increases[6-7] with time initially, reaches a peak, and finally decreases as time increases.

In accordance with the linear theory[30] of new phase formation, temporal variation of scattering function $S(q,t)$ is given by:

$$S(q,t) = S(q,0) \exp[2t\alpha(q)]$$

where $\alpha(q)$ is the time-independent proportionality constant.
It has been observed that $\alpha(q)$ does not behave time independent for all the hydrating specimen, discussed in the present work, even at the initial stage of the measurements, indicating inadequacy of the linear theory to comprehend the observations of the present set of measurements.
In order to examine the scattering function kinetics in the light of dynamical scaling phenomenon, based on nonlinear theories[1], of new phase formation, the normalized scaling function $\widetilde{F}[qL(t)] = S(q,t)[L(t)]^{-D_m} / \sum q^{D_m} S(q,t) \delta q$ has been calculated where $L(t) = \sqrt{\kappa(t)}$ and $\delta q$ is the experimental q increment.
It is important to note that for a mass fractal object, surface area also scales[31] as $r^{D_m}$ for a spherical surface of radius r. The variations of $\widetilde{F}[qL(t)]$ with $qL(t)$ for water mixture hydrating $C_3S$ with w/c=0.3 and w/c=0.5 are shown in Fig. 3. Different curves in each frame depict the variation of $\widetilde{F}[qL(t)]$ with $qL(t)$ for different recorded times. It is evident from the figure that the scaling functions are not strictly time independent indicating poor agreement with the scaling hypothesis. These results are in sharp contrast to those observed[6-7] in the case of light water hydrating specimens of silicates. In the case of hydrating silicates, good agreement with scaling hypothesis has been observed for a wide range of w/c values. It has also been observed that scaling phenomenon is also not operative for all the water mixture hydrating $C_3S$ specimens under investigation for $L(t)=[q_1(t)]^{-1}$, where $q_1(t)$ is the first moment of the scattering function $S(q,t)$.

It remains to be understood as to why scaling is observed in hydration of silicates, tricalcium silicate and dicalcium silicate, with pure $H_2O$ and not with either pure $D_2O$ or with mixture of $H_2O$ and $D_2O$. As it has been observed[9] that when silicates are mixed with $H_2O$, water rich C-S-H colloidal sol is formed initially. It is for this very reason the temporal evolution of mesoscopic structure of C-S-H gel will not be influenced by the microscopic structure of the silicate or on its particle size distribution although initial dissolution rate may depend on microscopic structure and on particle size distribution. The morphology of the $H_2O$ hydrating colloidal particle is that of a mass fractal with fractal dimension increasing with time initially and reaching a plateau.

For hydration with $D_2O$, the sol changes[6-7] topographically with time, unlike in the case of hydration with $H_2O$. In the beginning, the sol is ramified throughout the volume but the degree of ramification decreases with time. Subsequently, the mass transforms into objects with uniform internal density of un-ramified core with ramified surface showing self-similarity. Later, the ramified surface grows into ramified volume during hydration - the degree of ramification increasing with time. This topographical change of the hydrating mass as a function of time is one of the plausible reasons why scaling is not observed in the case of hydration of silicates with $D_2O$. Further, the formation rates of C-S-D and $Ca(OD)_2$ vary differently with time for hydration with $D_2O$. As a consequence of that the ratio of linear sizes of $Ca(OD)_2$ crystallite and C-S-D phase vary with time. This is another reason why scaling is not observed for hydration with $D_2O$. However, for hydration with water mixture, the sol does not change topographically with time. It remains mass fractal throughout with marginal variation of fractal dimension. So temporally varying formation rates leading to different growth kinetics of radii of C-S-H, C-S-D, $Ca(OH)_2$ and $Ca(OD)_2$ is a plausible reason why scaling is not observed in the case of hydration with water mixture. It is noteworthy that also both for light and heavy water hydration of sulphates, time evolution of the scattering functions do not exhibit scaling phenomenon[1] for a characteristic length with any possible measure although there is no topographical change of the hydrating mass as a function of time. Hydrating mass remains mass fractal throughout.

## IV. CONCLUSIONS

Role of hydrogen bond is crucial to understanding how cement, a crystalline powder, on mixing with liquid water turns into a monolithic mass, mass of structure in a one-piece form, of high compressive strength. Using pure light water, pure heavy water and mixture of the two as the hydration medium is one of the experimental ways to tune hydrogen bond effect. In the present work, temporal evolution of mesoscopic structure and hydration kinetics of cement with mixture of light and heavy water have been investigated. These experimental observations have been compared with the corresponding observations on the hydration of silicates with pure light or heavy water. It has been observed that dynamics of hydration of cement with mixture of light and heavy water is incomprehensible in terms of observed hydration dynamics of cements with pure light or heavy water.

Unlike in the case with heavy water and as in the case with light water, hydrating mass remains mass fractal throughout for hydration of silicates with water mixture – implying no topographical change of the hydrating mass as a function of time. Unlike in the case with light water, mass fractal dimension does not grow linearly with time initially – implying non dominant role of diffusion limited cluster aggregation mechanism for hydration of silicates with water mixture. But as in the case with heavy water and unlike in the case with light water, scaling phenomenon has not been observed for hydration of silicates with water mixture – implying non-synchronous formation of $Ca(OH)_2$ / $Ca(OD)_2$ and C-S-H / C-S-D in this period of hydration. As hydration of cement with light water exhibits dynamical scaling phenomenon, ratio of linear sizes of $Ca(OH)_2$ crystallite and C-S-H phase must be constant with time indicating the synchronous formation of $Ca(OH)_2$ and C-S-H in this period of hydration. The dynamics of evolution of the mesoscopic structure of hydration of cement has been observed to be nonlinear also in regard to the composition of hydration medium.

**Acknowledgement**

We are grateful to T. Mazumder and J. Bahadur for their help in making some figures.

**Figure Captions**

Fig. 1. A schematic representation of dynamical scaling. Individual inhomogeneities are scaled up by same time dependent scale factor L(t) implying some communication mechanism between the individual inhomogeneities.

Fig.2. Time evolution of scattering function S(q,t) for water mixture hydrating $C_3S$ with water-to-cement mass ratio (w/c) = 0.3 and 0.5, respectively. The inset shows the time evolution of S(q,t) at some specified values of q as specified therein. Statistical error bars are smaller than the respective symbol sizes.

Fig. 3. Scaled scattering function $\tilde{F}$[qL(t)] for water mixture hydrating $C_3S$ with water-to-cement mass ratio (w/c) = 0.3 and 0.5, respectively. The inset shows the time evolution of $\kappa(t)$ and $\eta(t)$, associated with power-law scattering. Statistical error bars are smaller than the respective symbol sizes. The solid lines are only guides to the eye.

Fig. 4. Model calculations depicting fluctuations of S(q,t) as a function q and volume fraction $\varphi$ of a fragmented inhomogeneity, for splitting of a spherical inhomogeneity into two spherical inhomogeneities. Volume fraction of one splitted inhomogeneity being $\varphi$.

Fig. 5. Model calculations depicting fluctuations of S(q,t) as a function of volume fraction and scattering length fraction at some specified q-values, for splitting of a spherical inhomogeneity into two spherical inhomogeneities. Volume fraction of one splitted inhomogeneity is φ with scattering length fraction φ$_\zeta$.

## Appendix

**Model Calculation:** Dealing with effect of changing coherence characteristic of inhomogeneties on S(q,t):

Let us consider a model system with one inhomogeneity of volume unity with total scattering length ζ at time t. At time t+δ, the inhomogeneity transforms into two spherical inhomogeneities with total scattering length (ζ-α) and α, respectively following the conservation of total scattering length and volume. There exists two possibilities in such situation.

(i) The scattering length densities of the fragmented spherical inhomogeneities remaining same:

Fig. 4 depicts the variation of $\frac{S(q,t+\delta)-S(q,t)}{S(q,t)}$ with q and volume fraction φ (=α/ζ) of one fragmented inhomogeneity. Variations are obviously symmetric about φ=0.5. The oscillatory nature of $\frac{S(q,t+\delta)-S(q,t)}{S(q,t)}$ is significant at q > (π/Diameter of sphere of volume unity) and significance of oscillation increases with increase of φ i.e., with increase of significance of fragmentation.

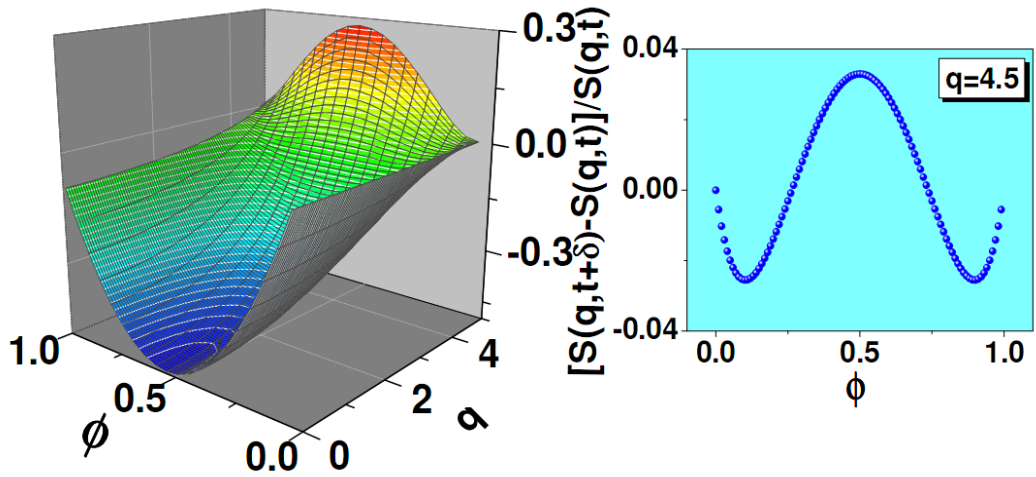

*Fig. 4. Model calculations depicting fluctuations of S(q,t) as a function q and volume fraction φ of a fragmented inhomogeneity, for splitting of a spherical inhomogeneity into two spherical inhomogeneities. Volume fraction of one splitted inhomogeneity being φ.*

(ii) Fragmented spherical inhomogeneities with varying scattering length densities:

Fig. 5 depicts the variation of $\frac{S(q,t+\delta)-S(q,t)}{S(q,t)}$ with q and volume fraction φ of one fragmented inhomogeneity with scattering length fraction $\varphi_\zeta$ (=α/ζ). At q→0, variations are parabolic and symmetric about $\varphi_\zeta$ =0.5. Symmetry breaks with increasing q. Results will be reversed when two or more inhomogeneities join together to form a coherent mass. A contiguous domain having uniform chemical composition and hence uniform scattering length density is termed as coherent mass. Composition modulation leads to incoherence. Scattering centres within a coherent mass scatter coherently.

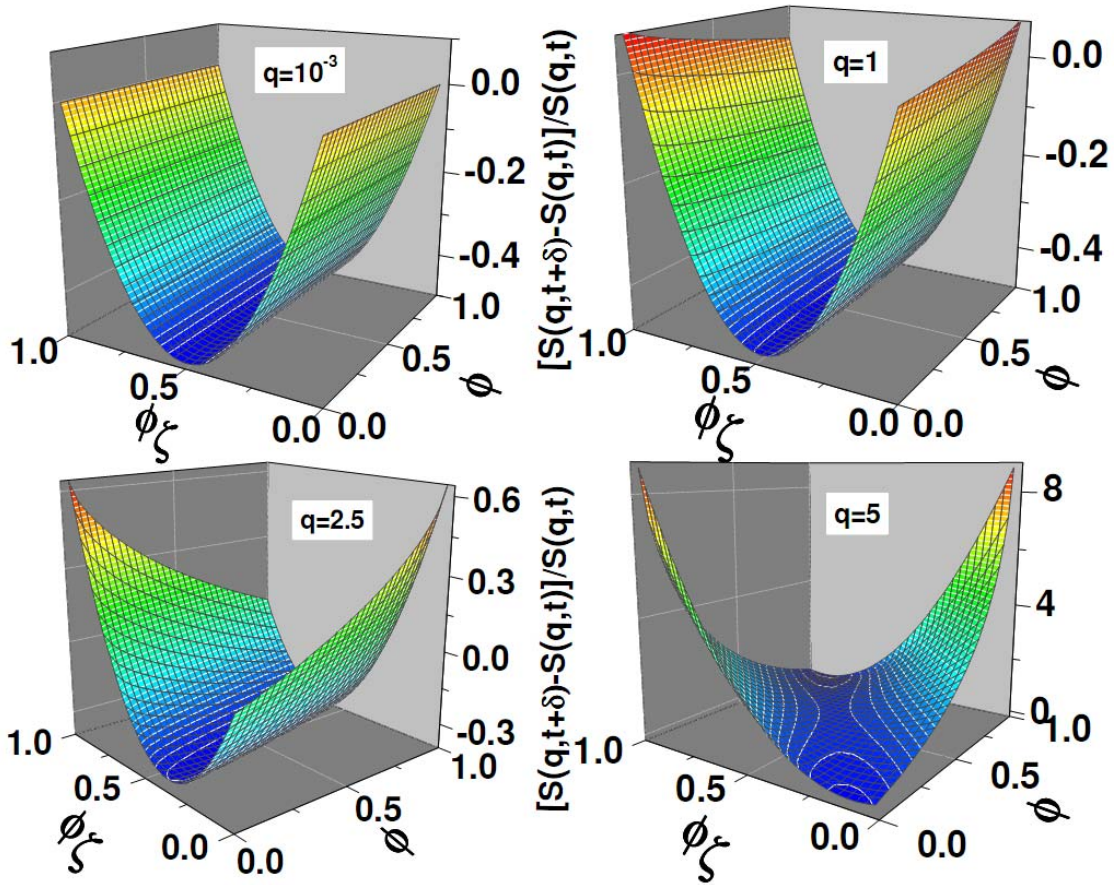

*Fig. 5. Model calculations depicting fluctuations of S(q,t) as a function of volume fraction and scattering length fraction at some specified q-values, for splitting of a spherical inhomogeneity into two spherical inhomogeneities. Volume fraction of one splitted inhomogeneity is φ with scattering length fraction $\varphi_\zeta$.*